\documentclass[aps,prd,twocolumn,nofootinbib]{revtex4-2}

\usepackage{booktabs,makecell,multirow}

\usepackage{graphicx}
\usepackage{amsmath}
\usepackage{amssymb}
\usepackage{bm}
\usepackage{bbm}
\usepackage{color}
\usepackage{slashed}
\usepackage{soul}
\usepackage{ulem}
\usepackage{tabularx}
\usepackage{caption}
\usepackage{subcaption}
\usepackage{indentfirst}
\usepackage{siunitx}

\usepackage[colorlinks=true, linkcolor=blue, citecolor=blue, urlcolor=blue]{hyperref}

\newcommand*{\RMN}[1]{\uppercase\expandafter{\romannumeral#1}}

\begin{document}
\title{Gravitational Waves from Higgs Preheating after Inflaton $Z_2$-Symmetry Breaking}

\author{Hua Zhou$^1$}
\email{zhouhua@swust.edu.cn}
\author{Qing Yu$^1$}
\email{yuq@swust.edu.cn}
\author{Wei Cheng$^2$}
\email{chengwei@cqupt.edu.cn}
\author{Ruo-Peng Zhang$^3$}
\email{zrp@cqvie.edu.cn}

\affiliation{$^1$ School of Mathematics and Physics, Southwest University of Science and Technology, Mianyang 621010, P.R. China\\
$^2$ School of Science, Chongqing University of Posts and Telecommunications, Chongqing 400065, P.R. China\\ 
$^3$ Chongqing Vocational Institute of Engineering, Big Data and Internet of Things School, Chongqing, 402260, P.R. China}

\date{\today}

\begin{abstract}
In this paper, nonperturbative lattice simulations are used to study Higgs preheating and the associated gravitational wave (GW) background after the inflaton $Z_2$ symmetry is broken during inflation. This symmetry breaking generates both trilinear and quartic inflaton-Higgs interactions during preheating. The quartic inflaton-Higgs coupling is characterized by $q_{\phi h}\equiv \lambda_{\phi h}/\lambda_\phi$, while the trilinear interaction enters jointly through $q_{\phi h}$ and $q_\epsilon\equiv m_\phi/(\sqrt{\lambda_\phi}\phi_0)$. The Higgs self-coupling parameter $q_h\equiv \lambda_h/\lambda_\phi$ determines the onset of backreaction through the effective mass induced by Higgs self-interactions. Our simulations show that efficient preheating requires both a sufficiently broad resonance band and delayed backreaction. For $\lambda_\phi=10^{-13}$, the viable parameter region is approximately $q_{\phi h}>10$, $q_h<10^3$, and $q_\epsilon<10^{-5}$. Smaller $q_\epsilon$ keeps the system in a quartic-dominated regime and suppresses the rapid drift of resonance bands, while smaller $q_h$ delays the end of preheating by weakening self-interaction-induced backreaction. The amplified Higgs inhomogeneities source GW through the transverse-traceless part of the anisotropic stress tensor. The lattice results show that the GW spectrum grows rapidly during parametric resonance, broadens through rescattering, and saturates in the nonlinear stage. At late times, the spectrum develops a broad peak with amplitude $\Omega_{\rm gw}\sim10^{-6}$ at production. After redshifting to the present day, the peak frequency is $f\sim10^9\,{\rm Hz}$ with present-day amplitude $\Omega_{\rm gw,0} h^2 \sim 10^{-10}$. These results suggest that high-frequency GW from Higgs preheating may be detectable by future resonant-cavity detectors.
\end{abstract}

\maketitle

\section{Introduction}
Inflation provides a natural explanation for the horizon, flatness, and magnetic monopole problems in early cosmology, while generating primordial density perturbations that seed the large-scale structure of the Universe~\cite{Starobinsky:1980te, Guth:1980zm, Sato:1981qmu, Guth:1982ec}. At present, inflation is strongly supported by high-precision measurements of the Cosmic Microwave Background (CMB) and has become one of the central pillars of the standard cosmological model~\cite{Fixsen:1996nj, Hanany:2000qf, Goldstein:2002gf, Dickinson:2004yr, SDSS:2005xqv, Sievers:2005gj, Reichardt:2008ay, QUaD:2009aub, Reid:2009xm, Chiang:2009xsa, WMAP:2010qai}. A wide variety of inflationary models have been proposed, including natural inflation, high-curvature inflation, and chaotic inflation~\cite{Freese:1990rb, Linde:1983gd, Zhou:2022ovp}. Meanwhile, increasingly precise cosmological observations have imposed stringent constraints on the parameter space of these models. In particular, measurements of primordial perturbations in the CMB provide tight constraints on the scalar spectral index $n_s$ and the tensor-to-scalar ratio $r$~\cite{Planck:2018jri, BICEP:2021xfz}, significantly narrowing the range of viable scenarios. Among these, models involving the Higgs sector during inflation or reheating are especially attractive~\cite{Futamase:1987ua, Cervantes-Cota:1995ehs, Bezrukov:2007ep}, as they establish a direct link between early-Universe cosmology and particle physics.

After inflation ends when the slow-roll conditions are violated ($\epsilon \approx 1$), the inflaton field $\phi$ begins to oscillate coherently around the minimum of its effective potential. Through reheating, the energy stored in the inflaton is eventually transferred into a thermal bath of relativistic particles. This process typically proceeds through three stages: parametric resonance, particle decay, and thermalization. The initial stage, known as preheating, is characterized by explosive particle production via non-perturbative parametric resonance~\cite{Traschen:1990sw, Shtanov:1994ce, Kofman:1994rk}. During preheating, energy is rapidly transferred from the inflaton field to coupled scalar or vector fields~\cite{Kofman:1994rk, Greene:1997ge, Khlebnikov:1997di, Bassett:1998wg, Easther:1999ws, Finelli:2001db, Podolsky:2005bw, Easther:2006gt}, in a highly non-equilibrium and non-perturbative manner~\cite{Dolgov:1982th, Abbott:1982hn}. Since parametric resonance acts only in specific momentum bands during preheating, the Fourier modes of inflaton quanta or scalar matter fields grow exponentially only within those bands. This selective growth leads to pronounced, time-varying density inhomogeneities in configuration space, which manifest as large anisotropic stresses. These field inhomogeneities efficiently act as sources for the gravitational wave (GW) background in the preheating; thus, the nonlinear dynamics during the preheating stage can give rise to a notable GW background~\cite{Kofman:1994rk, Khlebnikov:1997di, Easther:2006gt, Dufaux:2007pt, Adshead:2024ykw, Figueroa:2016ojl, Tranberg:2017lrx}.

The GW generated during the reheating epoch redshift as a radiation-like fluid during the subsequent cosmic expansion, remaining essentially decoupled from other energy-matter constituents. Crucially, unlike the GW from tensor perturbations sourced by vacuum fluctuations during inflation, the amplitude of preheating-generated GW is decoupled from the inflation energy scale, which determines their present-day peak density ranges from: $\Omega_{\rm gw}h^{2}\sim10^{-8}$ around the frequencies $f \sim 10^{8}$ Hz. Furthermore, with the enhanced understanding of the preheating dynamics theory and the detection range of the next-generation GW observatory approaching the preheating GW sensitivities~\cite{Yagi:2011wg, Sato:2017dkf, Reitze:2019iox, Sesana:2019vho, ET:2019dnz, Blas:2021mqw}. Therefore, conducting in-depth research on the GW generated during the preheating stage plays an important role in extracting the physical information of the universe at this stage.

Ref.~\cite{Khlebnikov:1997di} first revealed the GW background generated through preheating within quartic inflation models. Employing Weinberg's flat-spacetime formula, they calculated the first time that the gravitational wave energy per unit solid angle, identifying a background peak at $f\sim 10^{8}$ Hz. Ref.~\cite{Garcia-Bellido:2007nns} investigated hybrid inflation models using a flat-space numerical simulation that simultaneously evolved scalar fields and metric perturbations. The Ref.~\cite{Easther:2006vd} uses LatticeEasy~\cite{Felder:2000hq} to simulate scalar field dynamics in Friedmann-Robertson-Walker spacetime. Through coupling LatticeEasy with a metric perturbation integrator-while implementing cosmological expansion corrections, systematically studied hybrid inflation scenarios, discovering scale-invariant background amplitude. The Ref.~\cite{Garcia-Bellido:2007fiu} also studied the GW Background from Reheating after Hybrid Inflation using LatticeEasy with a metric configuration-space integrator within expanding spacetime. Meanwhile, Ref.~\cite{Dufaux:2007pt} established a method based on Green's function for studying the generation of gravitational waves from preheating after inflation.

Previous studies of Higgs preheating have shown that parametric resonance plays a central role in the post-inflationary dynamics, and that the efficiency of preheating depends sensitively on the interactions between the inflaton and the Higgs field. In the present work, we consider a two-field Higgs-preheating model in a spatially flat FRW background, in which an inflaton field $\phi$ is coupled to the Higgs field. The scalar potential is assumed to respect a $Z_2$ symmetry, $\phi \rightarrow -\phi$. After inflation, the inflaton oscillates around its true vacuum expectation value, and expanding the potential around the true vacuum naturally generates both trilinear and quartic interaction terms relevant for reheating. These interactions control the nonperturbative transfer of energy from the inflaton condensate to the Higgs sector. 

The dynamics of this system are mainly governed by three dimensionless parameters: the dimensionless Higgs self-coupling parameter $q_h=\lambda_h/\lambda_\phi$, the dimensionless inflaton–Higgs coupling parameter $q_{\phi h}=\lambda_{\phi h}/\lambda_\phi$, and the dimensionless mass-scale parameter $q_\epsilon = m_\phi/(\sqrt{\lambda_\phi}\phi_0)$. Physically, $q_{\phi h}$ controls the strength of the resonant excitation of Higgs modes, $q_\epsilon$ determines the stability of the resonance bands in momentum space, and $q_h$ regulates the onset of backreaction through the Higgs effective mass induced by self-interactions. The interplay among these parameters determines whether preheating can proceed efficiently, how long the resonance can be sustained, and how strongly nonlinear effects suppress the energy transfer.

In this paper, we perform a nonlinear lattice study of Higgs preheating with both trilinear and quartic inflaton--Higgs interactions, with particular emphasis on the role of Higgs self-interactions in regulating resonance, backreaction, and GW production. Using \texttt{CosmoLattice}, we determine the viable region of parameter space that allows efficient energy transfer from the inflaton to the Higgs sector, analyze the evolution of the Higgs power spectrum during preheating, and compute the corresponding GW energy-density spectrum. We then redshift the lattice-generated GW signal to the present epoch and compare it with existing cosmological bounds and projected sensitivities of high-frequency GW experiments. 

The remainder of this paper is organized as follows. In Sec.~\ref{sec2}, we introduce the Higgs preheating model, discuss the relevant constraints on the parameter space, and analyze the efficiency of energy transfer in lattice simulations. In Sec.~~\ref{sec3}, we investigate the production of gravitational waves, present the resulting spectra, and translate them into present-day observables. Finally, Sec.~~\ref{sec4} summarizes our results.

\section{Higgs preheating}
\label{sec2}
\subsection{Higgs potential}
In this work, we will consider a preheating model that couples an inflaton field $\phi$ with a Higgs field H. In the context of a spatially flat Friedmann-Robertson-Walker (FRW) universe, the relevant Lagrangian is expressed as:
\begin{eqnarray}
\mathcal{L}&=&\frac{1}{2}\partial_{\mu} \phi \partial^{\mu}\phi + (D_\mu {\rm H})^{\dagger}(D^{\mu} {\rm H}) - V(\phi,{\rm H}),
\end{eqnarray}
where $\rm H$ is the Higgs doublet
\begin{eqnarray}
\rm H &=& \frac{1}{\sqrt{2}} 
\begin{pmatrix} 
h_1 + i h_2 \\ 
h_3 + i h_4 
\end{pmatrix},
\end{eqnarray}
$h_j$ is the real scalar field and $j=\{1,2,3,4\}$. The potential is invariant under a $Z_2$ symmetry $\phi \to -\phi$ and can be expressed as follows
\begin{eqnarray}
V(\phi, {\rm H}) &=& \frac{\lambda_{\phi}}{4}\phi^{4} - \frac{1}{2}m_{\phi}^2\phi^2 + \frac{\lambda_{\phi h}}{2}\phi^{2} ({\rm H}^\dagger {\rm H}) + \lambda_{h} ({\rm H}^\dagger {\rm H})^{2}.\nonumber\\
\end{eqnarray}
During inflation, the dynamics is dominated by the inflaton field, while the Higgs field remains close to its vacuum state and plays a negligible role, 
the effective potential reduces to \footnote{We emphasize that the potential considered in this work is employed primarily as an effective description of the post-inflationary oscillatory stage. Its large field quartic extrapolation predicts $n_s \simeq 1-3/(N_{\ast}+1)$ and $r\simeq 16/(N_{\ast}+1)$, which are incompatible with current CMB constraints for $N_{\ast}\simeq50-60$~\cite{Planck:2018jri,BICEP:2021xfz}. A nonminimal coupling $\xi_{\phi}\phi^{2}R/2$ can flatten the potential and improve CMB compatibility, but also alters the canonical normalization, initial conditions, and possibly preheating dynamics~\cite{Bezrukov:2007ep,Garcia-Bellido:2008ycs,Ema:2016dny}. Thus, our lattice results are conditional on the adopted effective post-inflationary setup.}
\begin{eqnarray}
V(\phi) &=& \frac{\lambda_\phi}{4}\phi^4 -\frac{1}{2}m^{2}_{\phi}\phi^{2}.
\label{eq4}
\end{eqnarray}
Assuming that $Z_2$ symmetry is spontaneously broken, leading to a non-zero vacuum expectation value
\begin{eqnarray}
\langle \phi \rangle & = & v = \frac{m_\phi}{\sqrt{\lambda_\phi}}.
\end{eqnarray}
To study the reheating dynamics, we expand the inflaton field around the true vacuum. Meanwhile, in the unitary gauge, the Higgs doublet reduces to a single real scalar field,
\begin{eqnarray}
{\rm H} = \frac{1}{\sqrt{2}}
\begin{pmatrix}
0 \\ h
\end{pmatrix},
\quad \Rightarrow \quad {\rm H}^\dagger {\rm H} = \frac{1}{2} h^2.
\end{eqnarray}
In the present work, we focus on the scalar-sector dynamics during preheating and neglect the dynamical evolution of gauge fields. Accordingly, in the lattice simulations the covariant derivative is effectively reduced to the ordinary derivative. Gauge-field contributions to the anisotropic stress tensor are therefore not included in the computation of the gravitational-wave source. In terms of the shifted field $\phi$, the scalar potential becomes
\begin{eqnarray}
V(\phi,h)&=&\frac{\lambda_{\phi}}{4}\phi^{4}+ m_{\phi} \lambda_{\phi}^{\frac{1}{2}}\phi^{3}+\frac{1}{4}\lambda_{\phi h}\phi^{2}{h}^{2}+m_{\phi}^{2}\phi^{2}\nonumber\\
&+&\frac{1}{2}m_{\phi}\lambda_{\phi}^{-\frac{1}{2}} \lambda_{\phi h}\phi{h}^{2}+\frac{1}{4}\lambda_{\phi}^{-1}\lambda_{\phi h}m^{2}_{\phi}{h}^{2} + \frac{1}{4}\lambda_{h}{h}^{4}.\nonumber\\
\end{eqnarray}

\subsection{Model constraints}
To ensure efficient energy transfer from the inflaton to the daughter fields and the generation of significant GW signals, the model parameters should be carefully controlled in three key aspects: the intensity of parametric resonance, the duration of the resonance, and the suppression of backreaction~\cite{Dufaux:2007pt, Figueroa:2017vfa, Fan:2021otj, Mansfield:2023sqp}. First, the parameter governing energy transfer is the inflaton–Higgs coupling parameter $q_{\phi h} = \lambda_{\phi h} / \lambda_{\phi}$, which dictates the magnitude of the Higgs field excitation within the broad resonance regime. It is well known that sufficiently large values of $q_{\phi h}$ can place the system in a broad-resonance regime. In quartic-type preheating, however, the detailed resonance structure depends nontrivially on the inflaton–Higgs coupling parameter, so the efficiency is not strictly monotonic in $q_{\phi h}$~\cite{Kofman:1994rk, Khlebnikov:1996wr}. In this case, even when accounting for the decay of field amplitudes due to cosmic expansion, the Higgs modes enter and remain within instability bands, leading to an exponential growth in energy density. 

The mass-scale parameter $q_\epsilon = m_{\phi} / (\sqrt{\lambda_{\phi}} \phi_0)$ (where $\phi_0$ denotes the initial amplitude of the inflaton field) measures how far the system is from the quartic-dominated. It also determines the dynamic change from quartic-driven to mass-dominated oscillations. For small $q_\epsilon$, the system remains approximately conformal, and the oscillation frequency is nearly constant in conformal time. Consequently, the resonance windows in momentum space stay stable, allowing for sustained energy transfer. Conversely, if $q_\epsilon$ is large, the mass term $m_{\phi}^2 \phi^2$ dominates the dynamics. In such mass driven preheating, the resonance windows drift rapidly as the field amplitude decays with the scale factor, causing the resonance to terminate prematurely. Therefore, to ensure sufficient energy transfer, a small $q_\epsilon$ is typically preferred to keep the system within the quartic-dominated regime~\cite{Greene:1997fu}. Finally, the Higgs self-coupling parameter $q_{h} = \lambda_{h} / \lambda_{\phi}$ controls the upper bound of energy transfer. Once the Higgs field is excited via resonance, its self-interaction induces an effective mass that alters the resonance phase conditions, triggering a backreaction that eventually terminates the process.

To maximize the energy transferred from the $\phi$ field to the $h$ field, the onset of backreaction must be delayed; a smaller $q_h$ allows the $h$ field to attain larger amplitudes. At the electroweak scale, LHC measurements of the Higgs mass imply $\lambda_h \approx 0.13$~\cite{CMS:2012qbp, ATLAS:2012yve}. However, within the experimentally allowed parameter space, the running of the Higgs self-coupling at higher energy scales remains only weakly constrained. Renormalization group evolution typically drives $\lambda_h$ to smaller values at high energies; concurrently, vacuum stability requires that $\lambda_h$ be positive, so $\lambda_h$ lies within the range $[0, 0.13]$~\cite{Elias-Miro:2011sqh}. In the subsequent numerical analysis, we fix $\lambda_{\phi} = 10^{-13}$ and vary the values of $q_{\phi h}$, $q_{h}$, and $q_\epsilon$. This approach allows us to systematically investigate how the strong backreaction induced by Higgs self-interactions affects the preheating dynamics and the resulting GW background.

\subsection{Preheating in the Lattice simulation}
This section investigates Higgs preheating following inflation. Driven by non-perturbative parametric resonance, this stage involves complex field interactions that transcend the scope of perturbative approaches. Consequently, we employ publicly available lattice code \texttt{CosmoLattice} to capture the non-linear evolution and backreaction effects that are inaccessible via the standard Boltzmann equation~\cite{Figueroa:2020rrl, Figueroa:2021yhd}. In the simulation of this article, we choose the simulation parameters such that the relevant physical scales are well separated from both the infrared (IR) and ultraviolet (UV) cutoffs. We employ a lattice size of $N=256^{3}$ \footnote{We have checked that increasing the lattice resolution (larger $N$) and varying the box size (smaller $k_{IR}$) do not lead to significant changes in the resulting spectra, confirming the robustness of our results.}. To ensure stable and accurate numerical evolution, we use $dt=0.01$ for the benchmark parameter and adopt smaller time steps at larger $q_{\phi h}$, with $dt/\delta x < 1/\sqrt{3}$ satisfied in all simulations \footnote{\interlinepenalty=10000 As $q_{\phi h}$ increases, the inflaton-induced contribution to the effective Higgs mass becomes larger, raising the characteristic frequency scale of the Higgs modes. We therefore reduce the time step until the energy-density evolution and field spectra become insensitive to the time step.}. The IR cutoff is defined as $k_{IR} = 2\pi/L = 0.75$, where $L$ denotes the comoving box side length, and this fixes the lattice spacing to $\delta x=L/N$. The momentum cutoff for initial fluctuations is set to $k_{cutoff}=4$, which restricts the initial excitation of field modes to momenta below this value in program units. The UV cutoff is determined by the grid discretization scheme. It is given by
\begin{eqnarray}
k_{UV} \sim \frac{\pi}{\delta x}=\frac{N}{2} k_{IR},
\end{eqnarray}
while the maximum momentum satisfies the relation 
\begin{eqnarray}
k_{max}&=&\sqrt{3}\frac{\pi}{\delta x},
\end{eqnarray}
This cutoff defines the maximum resolvable momentum in the simulation, ensuring numerical consistency. For the initial conditions, vacuum fluctuations are imposed on the scalar fields to mimic quantum fluctuations. These are implemented as Gaussian random fields in momentum space. Specifically, for scalar fields, the fluctuation amplitudes are sampled from a Rayleigh distribution to match quantum expectations, while the phases are drawn from a uniform distribution to ensure randomness and isotropy.

\begin{figure}[htbp]
\centering
\includegraphics[width=0.48\textwidth]{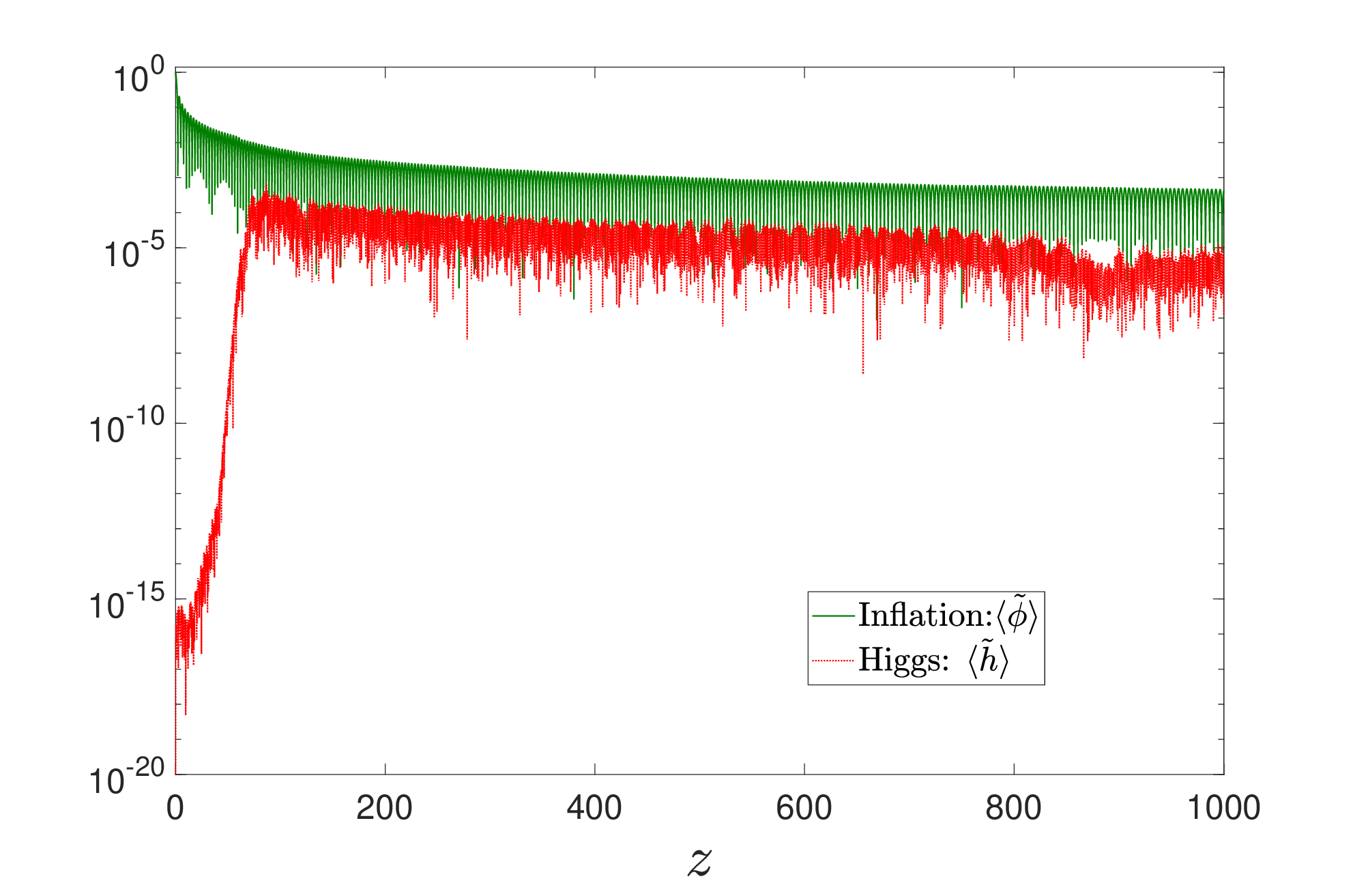}
\caption{Evolution of the Higgs and inflaton mean field values, where $\lambda_{\phi}=10^{-13}$, $q_{\phi h}=10^{2}$, $q_\epsilon=10^{-6}$, and $q_{h}=10^{2}$.}
\label{figfield}
\end{figure}

Firstly, the dynamic evolution of the absolute values of the mean Higgs field value $\langle \tilde h \rangle$ (red dotted line) and the inflaton field $\langle \tilde \phi \rangle$ (green solid line) is shown in Fig.\ref{figfield}. The horizontal axis represents the conformal time $z$. It is defined by $z = \sqrt{\lambda_{\phi}} \phi_{0} \int dt/ a(t)$, where $a(t)$ is the scale factor. The vertical axis indicates the inflaton field amplitude and the Higgs field amplitude. In the initial stage, the Higgs field undergoes rapid growth driven by parametric resonance; the mean value $\langle \tilde h \rangle$ increases by several orders of magnitude before reaching a non-linear saturation stage at an amplitude of approximately $\mathcal{O}(10^{-5})$. Meanwhile, the amplitude of the inflaton field decays, indicating the commencement of energy transfer from the inflaton to the sub-dominant Higgs sector. The fact that $\langle \tilde h \rangle$ remains smaller than the inflaton scale indicates that the zero mode of the Higgs field is not significantly amplified. Instead, the energy is predominantly transferred into inhomogeneous fluctuations with nonzero momentum. This process leads to the formation of spatially inhomogeneous structures in the Higgs field~\cite{Felder:2000hj, Felder:2001kt, Garcia-Bellido:2002fsq}, making this stage a significant potential source of GW. As the evolution enters the non-linear regime ($z \gtrsim 100$), the growth of the mean Higgs field saturates and begins to exhibit a slow suppression, a behavior attributed to the backreaction of produced particles and the scattering between different modes. The energy eventually cascades to smaller scales through turbulence; these dynamical evolutions are consistent with hybrid inflation models, where preheating is driven by strong interfield couplings~\cite{Garcia-Bellido:2007fiu}. A comprehensive analysis of the field dynamics is provided in Ref.~\cite{Garcia-Bellido:2002fsq}, the present work specifically focuses on the details of GW production during this epoch.

In realistic cosmological scenarios, most of the inflaton energy should be transferred to Standard Model radiation predominantly via the Higgs field. However, backreaction and rescattering can significantly suppress this process. Ref.~\cite{Lebedev:2021zdh} showed that, in the absence of a Higgs self-coupling ($\lambda_h \approx 0$), the Higgs field amplitude grows exponentially due to parametric resonance~\cite{Khlebnikov:1996mc, Prokopec:1996rr, Bernal:2018hjm}, which finally leads to a strong backreaction on the inflaton. However, the introduction of a non-zero $\lambda_h$ induces an additional effective mass term that dynamically shifts the position of the resonance bands. This mass-induced shift can either suppress or facilitate the resonance efficiency, thereby modulating the overall rate of energy transfer between the two fields. It should be noted that $\lambda_h > 0$ is required to ensure the stability of the Higgs potential at the electroweak scale. In the present setup, both trilinear and quartic inflaton--Higgs interactions arise naturally after expanding the $Z_2$-symmetric potential around the true inflaton vacuum. We therefore perform a systematic analysis of the preheating dynamics, focusing on the influence of the inflaton–Higgs coupling parameter $q_{\phi h}$, the mass-scale parameter $q_\epsilon$, and the Higgs self-coupling parameter $q_{h}$.

We next compute particle production using the adiabatic formalism developed in Ref.~\cite{Kofman:1997yn}. The analysis is performed in conformal time $\tau$, defined via $d\tau=dt/a(t)$, and introduce the canonically normalized field
\begin{eqnarray}
\tilde h(\tau,\mathbf{x}) \equiv a(\tau)  h(\tau,\mathbf{x}).
\end{eqnarray}
The field is subsequently decomposed into Fourier modes using comoving momentum $k$:
\begin{eqnarray}
\tilde h(\tau,\mathbf{x}) = \int \frac{d^3 \mathbf{ k}}{(2\pi)^3} \tilde h_k(\tau)  e^{i\mathbf{k}\cdot\mathbf{x}}.
\end{eqnarray}
Each mode satisfies the equation of motion
\begin{eqnarray}
\tilde h_k'' + \omega_k^2(\tau)  \tilde h_k = 0,
\end{eqnarray}
where the effective frequency is
\begin{eqnarray}
\omega_k^2(\tau)=k^{2} +a^{2}(\tau)m^{2}_{h}(\tau) -\frac{a''(\tau)}{a(\tau)},
\end{eqnarray}
and primes denote derivatives with respect to conformal time. The adiabatic particle occupation number is defined as
\begin{eqnarray}
n_k(\tau) = \frac{1}{2\omega_k} \left( |\tilde h_k'|^2 + \omega_k^2 |\tilde h_k|^2 \right) - \frac{1}{2}.
\end{eqnarray}
and the number density and energy density are given by
\begin{eqnarray}
n_h(\tau) &=& \frac{1}{(2 \pi)^3 a^3(\tau)} \int d^3 k  n_k(\tau),
\label{eq22}
\end{eqnarray}
\begin{eqnarray}
\rho_h(\tau) &=& \frac{1}{(2\pi)^3 a^4(\tau)} \int d^3 k  \omega_k(\tau)  n_k(\tau).
\label{eq23}
\end{eqnarray}
The effective Higgs mass is obtained from the scalar potential as
\begin{eqnarray}
m_h^2(\tau) = \frac{\partial^2 V}{\partial h^2}.
\end{eqnarray}
For the scalar potential considered above, this yields
\begin{eqnarray}
m_h^2(\tau) = \frac{1}{2}\lambda_{\phi h} \left(\phi + \frac{m_\phi}{\sqrt{\lambda_\phi}}\right)^2
+ 3\lambda_h h^2.
\end{eqnarray}
To capture the non-linear dynamics, we numerically evolve the classical equations of motion using the lattice code \texttt{CosmoLattice}, from which the macroscopic quantities defined in Eqs.~\ref{eq22} and~\ref{eq23} are evaluated. 

\subsection{Efficiency of inflaton-to-Higgs energy transfer}
\label{sec24}

This section investigates the efficiency of energy transfer from the inflaton to the Higgs field during preheating. After the end of inflation, the inflaton field oscillates around the minimum of its potential, leading to non-perturbative particle production via parametric resonance through its interaction with the Higgs field. We systematically explore the model parameter space to identify the conditions under which efficient preheating can be sustained. To isolate the role of the Higgs self-coupling parameter, we fix the inflaton–Higgs coupling parameter $q_{\phi h} = 10^{2}$ and the mass-scale parameter $q_\epsilon=10^{-6}$, and analyze the sensitivity of the energy transfer efficiency to variations in $q_h$. By varying $q_h$, we probe the impact of Higgs self-interactions on the nonlinear evolution, particularly the onset of backreaction and rescattering effects that can significantly alter the efficiency of resonance. This allows us to identify the region of parameter space where resonance is sustained long enough to enable efficient energy transfer, as well as the regimes in which strong self-interactions suppress particle production and prematurely terminate preheating.

\begin{figure*}[htbp]
     \centering
     \begin{subfigure}[b]{0.48\textwidth}
         \centering
         \includegraphics[width=\textwidth]{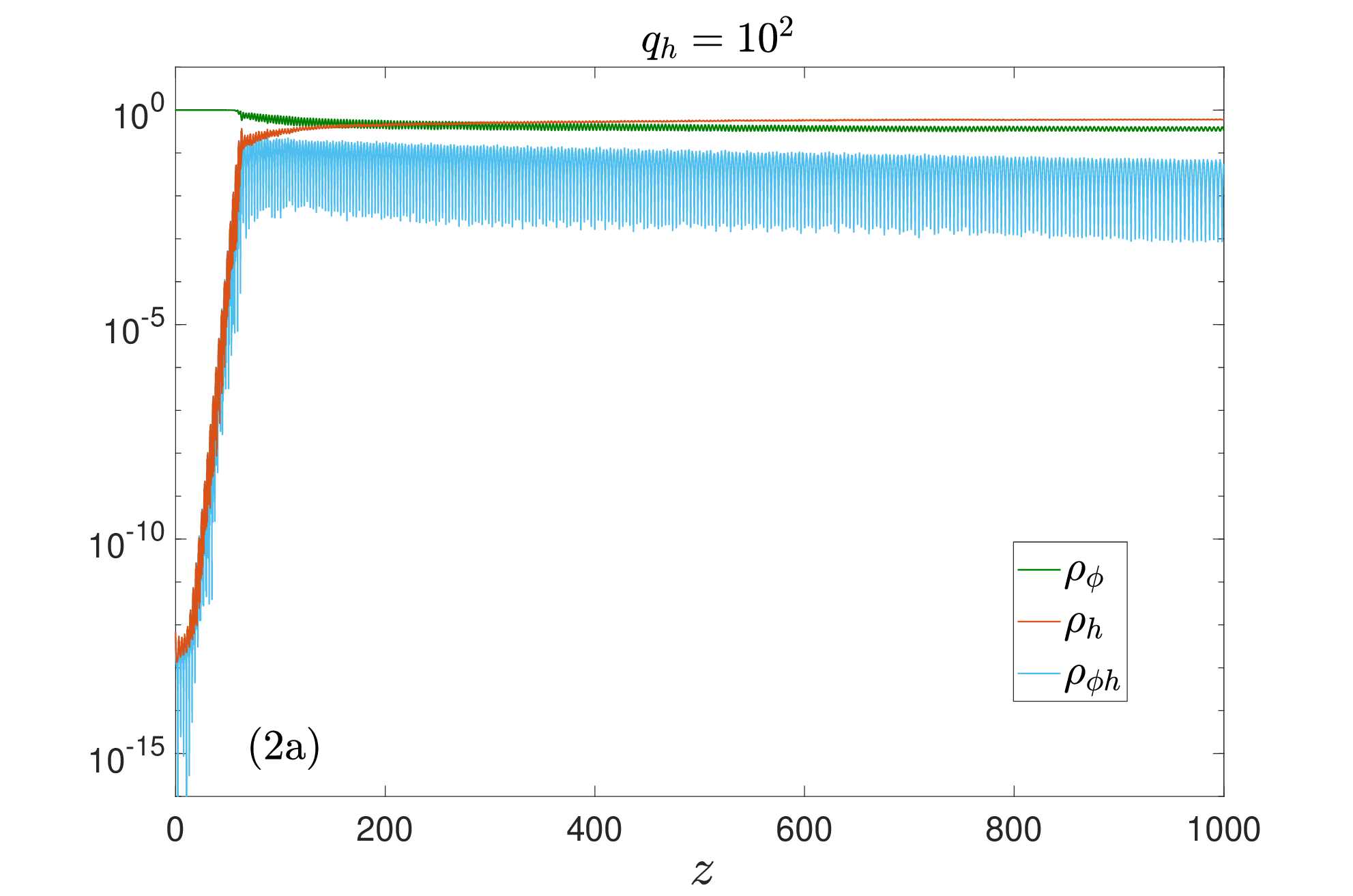}
          \caption{ }
         \label{figsub-a}
     \end{subfigure}
     \hfill
     \begin{subfigure}[b]{0.48\textwidth}
         \centering
         \includegraphics[width=\textwidth]{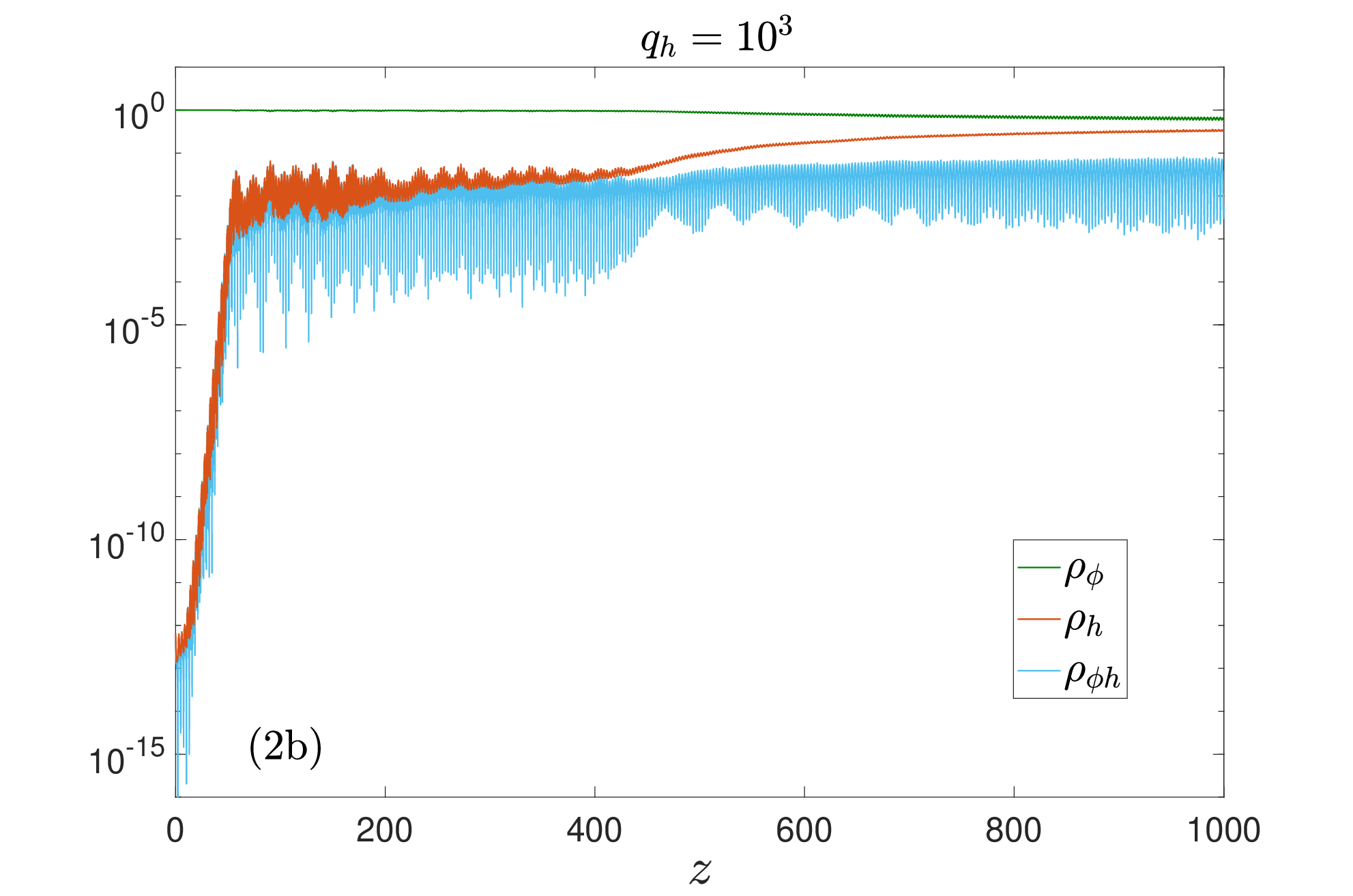}
          \caption{ }
         \label{figsub-b}
     \end{subfigure}
    \hfill
     \begin{subfigure}[b]{0.48\textwidth}
         \centering
         \includegraphics[width=\textwidth]{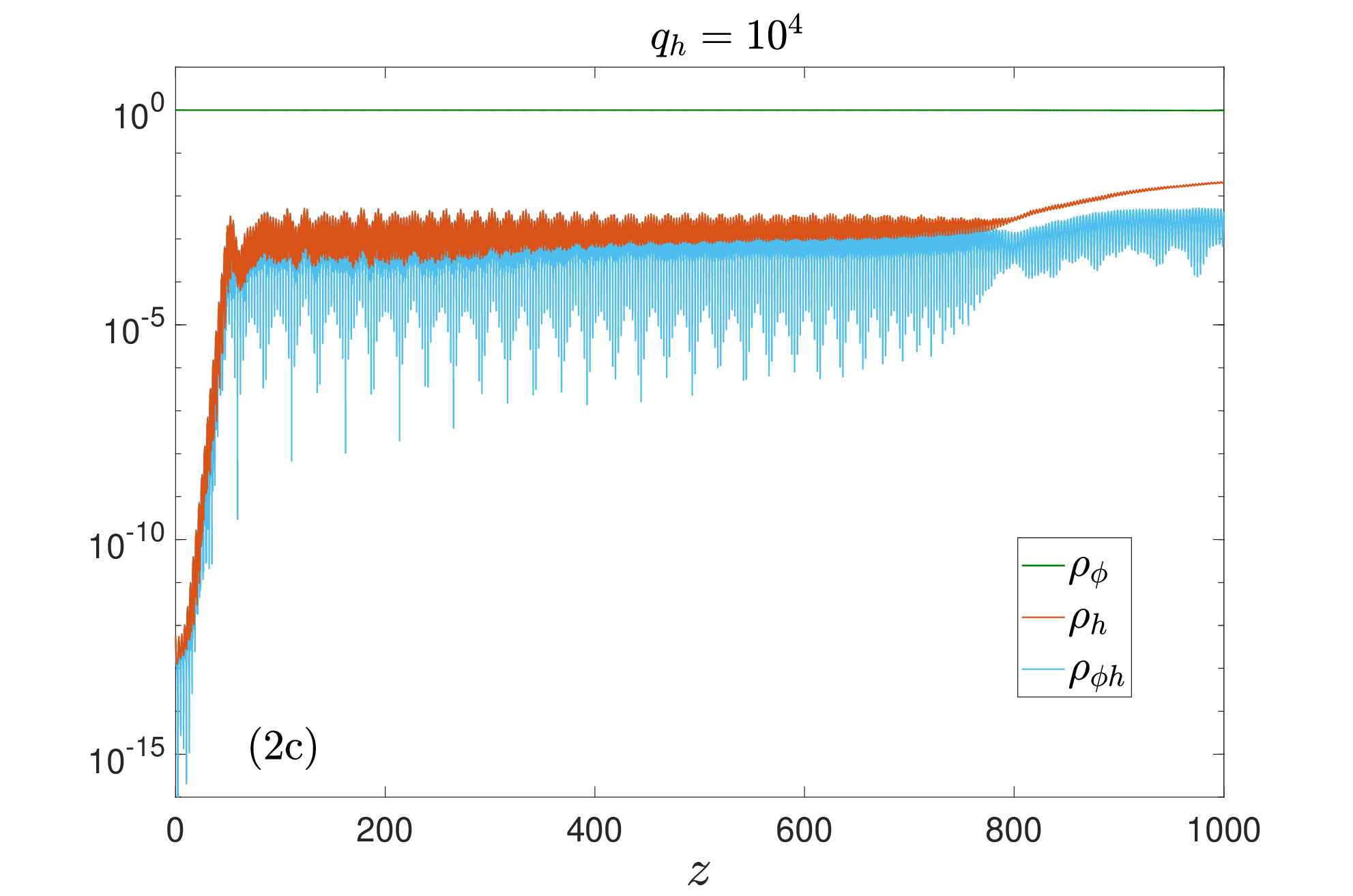}
          \caption{ }
         \label{figsub-c}
     \end{subfigure}
   \hfill
     \begin{subfigure}[b]{0.48\textwidth}
         \centering
         \includegraphics[width=\textwidth]{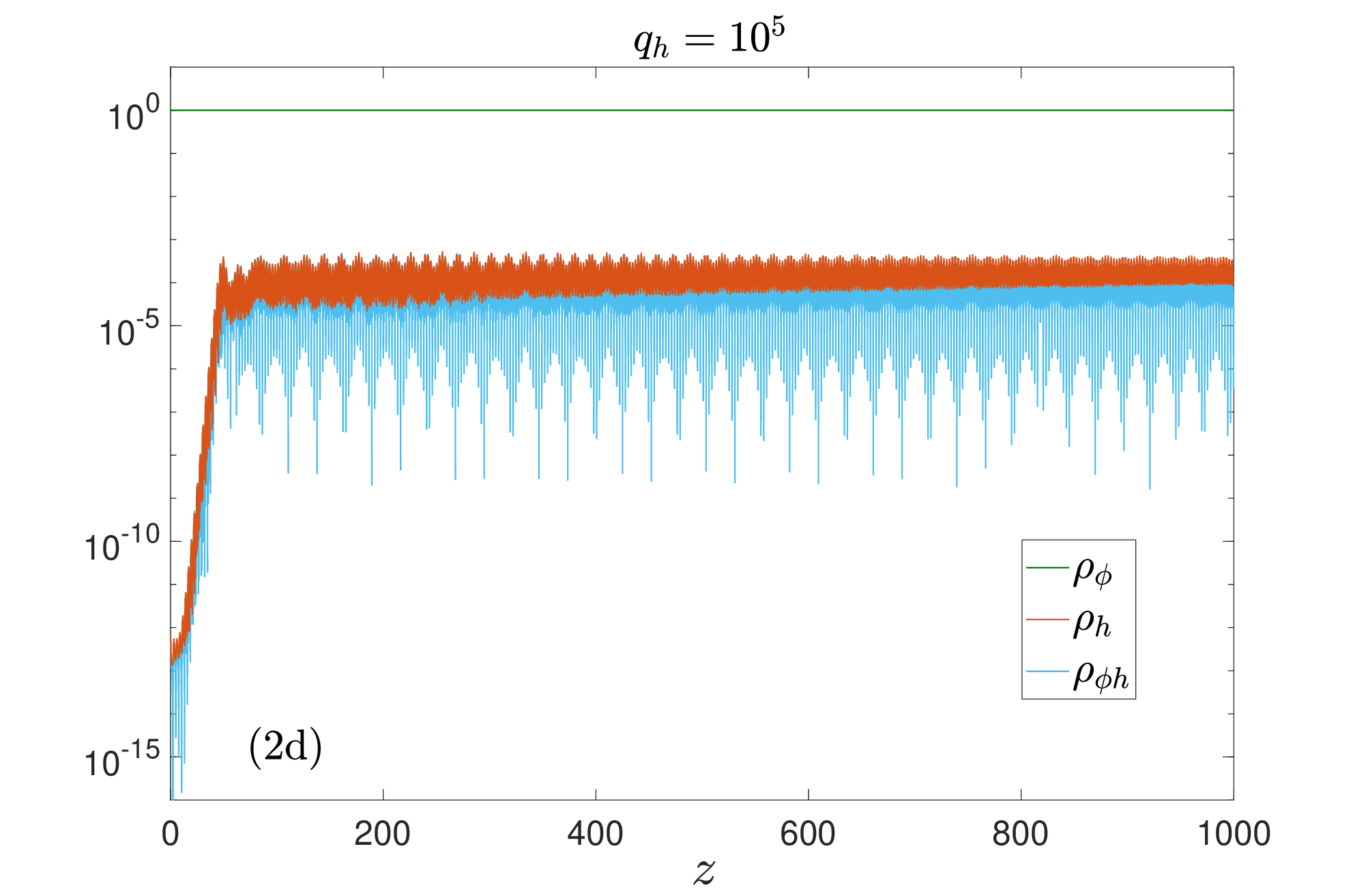}
          \caption{ }
         \label{figsub-d}
     \end{subfigure}
\caption{The evolutions of the energy density of Higgs field $\rho_h$, inflaton field $\rho_{\phi}$ and the interaction energy $\rho_{\phi h}$ as a function of the conformal time, where the energy has been normalized to the total energy density. The parameters are set to $\lambda_{\phi}=10^{-13}$, $q_{\phi h}=10^{2}$, $q_\epsilon=10^{-6}$.}
     \label{figenergy}
\end{figure*}

Fig.~\subref{figsub-a} illustrates the evolution of the normalized energy density components, as a function of the dimensionless conformal time $z$. When $q_h=10^{2}$, the interaction $\rho_{\phi h}$ (light blue line) and the Higgs energy density $\rho_h$ (the brown line) increase nearly exponentially, and their values rise rapidly from a very low level within a short period ($z \approx 50$), spanning several orders of magnitude, which is the signature of the explosive onset of parametric resonance instability. Concurrently, the inflaton energy density $\rho_{\phi}$ (the olive line) decays efficiently, dropping from an initial value of unity, and eventually the energy of $\rho_{\phi}$ is smaller than $\rho_h$. For $z \gtrsim 50$, all energy components stabilize: the ratio between $\rho_h$ and $\rho_{\phi}$ fluctuates slightly around 1, while $\rho_{\phi h}$ decays and remains at a very low value thereafter. 

This behavior is characteristic of parametric resonance during preheating. The exponential growth of $\rho_{h}$, together with the decay of $\rho_{\phi}$, indicates an efficient transfer of energy from the inflaton to the Higgs field. The strong oscillations in $\rho_h$ and $\rho_{\phi h}$ during the initial resonance phase reflect the spatiotemporal variation of the resonance parameter and the non-perturbative nature of particle production. Subsequently, backreaction effects from the produced particles suppress the resonance, leading the system toward equilibrium. At this stage, the negligible contribution of $\rho_{\phi h}$ indicates that the direct coupling energy between the fields becomes minimal, with most of the energy residing in the form of free particles associated with $\rho_{\phi}$ and $\rho_{h}$.

In contrast, when the Higgs self-coupling ratio $q_h$ approaches $q_h \approx 10^{3}$, the energy transfer becomes inefficient. As shown in Fig.~\subref{figsub-b}, the inflaton energy density $\rho_{\phi}$ remains dominant throughout the evolution, indicating that only a small fraction of energy is transferred to the Higgs sector. The intermediate case $q_h = 10^4$, shown in Fig.~\subref{figsub-c}, exhibits a partially suppressed energy transfer. This suppression of energy transfer is further exacerbated as the Higgs self-coupling ratio increases to $q_h=10^5$. In this case, as illustrated in Fig.~\subref{figsub-d}, the rapid onset of backreaction terminates the energy redistribution at an early stage. Consequently, both the Higgs energy density and the interaction energy remain strongly suppressed, exhibiting no significant growth. 

The behavior of these curves indicates that, although weak energy transfer does occur, the vast majority of the energy remains confined within the inflaton field. Therefore, preheating is inefficient in this parameter region. This can be traced back to the strong Higgs self-interaction, which induces substantial backreaction effects. These effects generate sizable effective mass corrections and enhance rescattering processes, thereby disrupting the resonance structure and suppressing non-perturbative particle production. As a result, the reheating process remains inefficient, and the universe remains dominated by the energy of the inflaton field. Overall, our analysis demonstrates that the efficiency of energy transfer is highly sensitive to the Higgs self-coupling parameter $q_h$. Increasing $q_h$ enhances backreaction, accelerates the growth of the Higgs effective mass, and modifies the resonance conditions, leading to a premature termination of the explosive particle production phase and, consequently, a reduced preheating efficiency. 

Meanwhile, we emphasize that $q_h$ should be understood as an effective high-scale Higgs self-coupling ratio during the preheating stage. Since we fix $\lambda_\phi=10^{-13}$, the condition $q_h<10^3$ corresponds to an effective $\lambda_h<10^{-10}$. Therefore, the efficient-preheating region implicitly assumes that the Higgs self-coupling is strongly suppressed at the high-energy scale. If the Standard Model value $\lambda_h\simeq0.13$ were directly used without such suppression, the corresponding $q_h$ would be much larger, and the resonance would be strongly affected by backreaction.

\begin{figure}[htbp]
\centering
\includegraphics[width=0.48\textwidth]{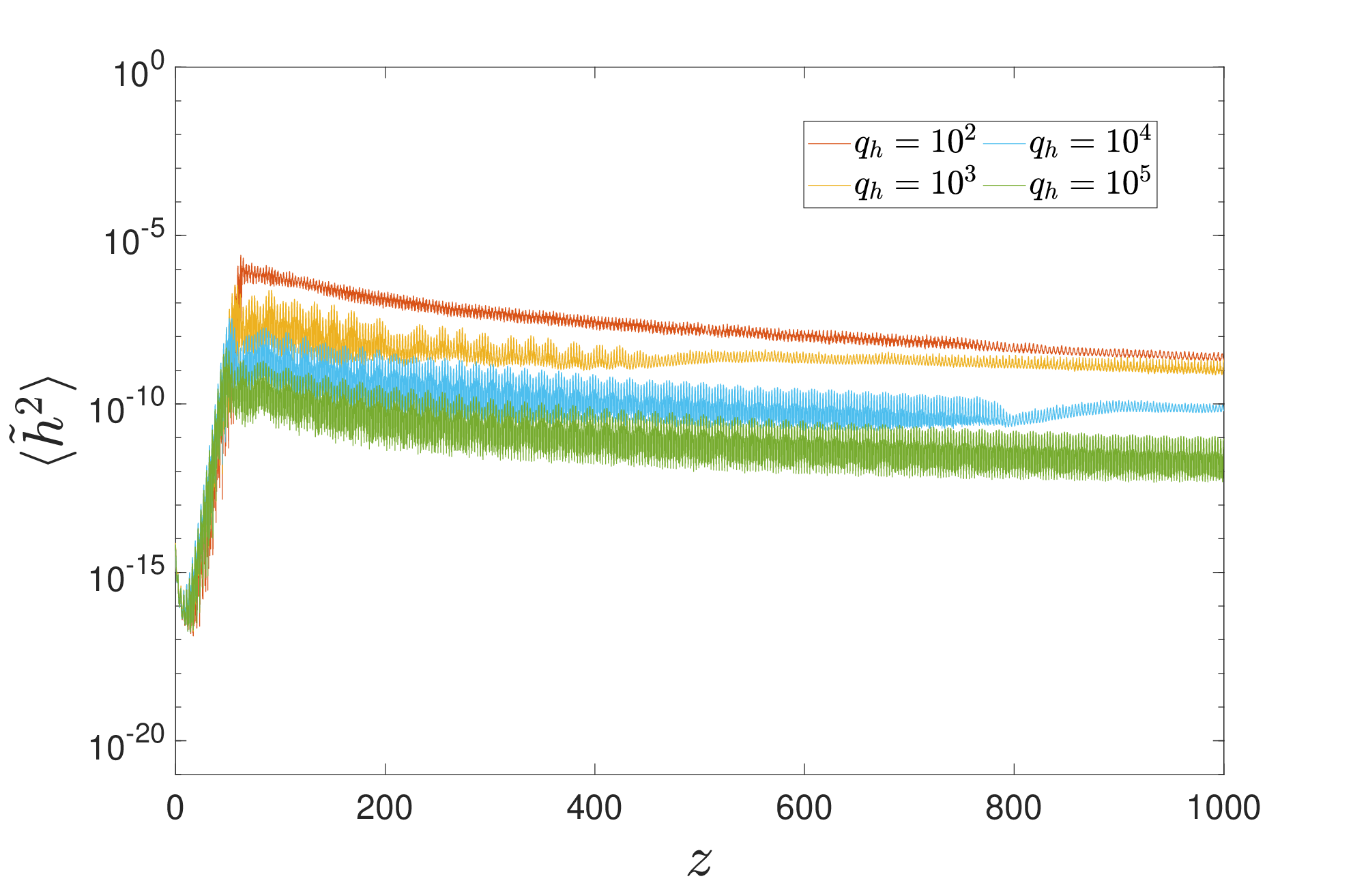}
\caption{The evolutions of the volume-averaged mean-square Higgs field \(\langle \tilde h^2 \rangle\) as a function of the dimensionless time $z$. The parameters are set to $\lambda_{\phi}=10^{-13}$, $q_{\phi h}=10^{2}$, $q_\epsilon=10^{-6}$.}
\label{figthreepanels}
\end{figure}

Then, Fig.~\ref{figthreepanels} shows the evolution of the volume-averaged mean-square Higgs field \(\langle \tilde h^2 \rangle\) for different $q_h$. At the initial stage, $\langle \tilde h^2 \rangle$ exhibits a rapid growth, indicating an efficient amplification of field fluctuations driven by parametric resonance. As the system evolves, the saturation value of $\langle \tilde h^2 \rangle$ does not increase with $q_h$. In particular, for $q_h = 10^{2}$, the final amplitude is several orders of magnitude larger than in cases with higher $q_h$. Specifically, the peak value reaches about $10^{-6}$ when $q_h = 10^{2}$, while it decreases to the order of $10^{-9}$ when $q_h$ increases to $10^{5}$. This shows that a stronger self-coupling does not always lead to more particle production. Instead, when $q_h$ is large enough, particle production is strongly suppressed.

This suppression can be understood as a consequence of enhanced backreaction effects. A larger Higgs self-coupling leads to a rapid growth of the effective mass of the Higgs field, which modifies the resonance structure and shifts the system out of the instability bands. At the same time, nonlinear rescattering becomes increasingly efficient, redistributing energy among modes and further disrupting coherent resonance. As a result, the resonance is terminated prematurely, leading to a lower saturation amplitude. These results are consistent with the behavior observed in the energy density evolution in Fig.~\ref{figenergy}. Efficient energy transfer and strong field amplification occur only within a narrow range of $q_h$, while larger values of $q_h$ suppress both the resonance and the energy transfer.

Furthermore, we extend our analysis to explore the viable parameter space associated with the inflaton-Higgs coupling parameter $q_{\phi h}$ and the mass-scale parameter $q_\epsilon$. The corresponding constraints are summarized in Table~\ref{tab1}. By fixing the inflaton self-coupling to $\lambda_{\phi} = 10^{-13}$, we systematically explore the region of parameter space compatible with efficient preheating. Numerical results suggest that efficient preheating is favored in the approximate range $q_{\phi h}>10$.\footnote{A possible upper bound on $q_{\phi h}$ cannot be excluded. However, no such bound is found within the parameter range explored in our simulations. Exploring larger $q_{\phi h}$ would require much finer temporal and spatial resolutions, beyond the reach of our current numerical setup.} Similarly, efficient preheating favors $q_\epsilon\lesssim10^{-5}$, since larger values lead to a rapid drift of the resonance bands and an earlier termination of particle production. These results demonstrate that efficient preheating occurs only within a restricted region of parameter space, where the resonance is sufficiently strong, while backreaction effects remain under control.

\begin{table*}[htbp]
    \centering
    \renewcommand{\arraystretch}{1.5}
    \begin{tabular}{lc}
        \toprule
        Parameter & Viable range for efficient preheating \\
        \midrule
        Higgs self-coupling parameter  ($q_h$) & $< 10^{3}$ \\
        inflaton–Higgs coupling parameter ($q_{\phi h}$) & $>10$ \\
        mass-scale parameter ($q_\epsilon$) & $< 10^{-5}$ \\
        \bottomrule
    \end{tabular}
    \caption{Viable parameter space for the Higgs self-coupling parameter, inflaton–Higgs coupling parameter, and mass-scale parameter obtained from numerical simulations, where we fix $\lambda_{\phi}=10^{-13}$.}
\label{tab1}
\end{table*}

\section{Production of the gravitational-wave background}
\label{sec3}
\subsection{Higgs dynamics and GW production}
In the following, we investigate the production of the gravitational-wave background during Higgs preheating by lattice simulations. In a spatially flat Friedmann-Robertson-Walker (FRW) universe, gravitational waves are described by the transverse-traceless (TT) component of metric perturbations~\cite{Dufaux:2007pt}, i.e.,
\begin{eqnarray}
ds^{2}&=&g_{\mu\nu}dx^{\mu}dx^{\nu}=-dt^{2} +a^{2}(t) (\delta_{ij}+h_{ij}) dx^{i}dx^{j}.\nonumber\\
\end{eqnarray}
The tensor perturbation $h_{ij}$ satisfies the conditions $\partial_{i}h_{ij}=h_{ii}=0$, and it corresponds to two independent tensor degrees of freedom. Its dynamics are governed by the equation of motion
\begin{eqnarray}
	\ddot{h}_{ij}+ 3H \dot{h}_{ij}-\frac{1}{a^{2}}\bigtriangledown^{2}h_{ij}&=& \frac{16\pi G}{a^2} \Pi^{TT}_{ij},
\label{eqf}
\end{eqnarray}
where $H(t)=\dot{a}(t)/a(t)$ is the Hubble parameter. In the absence of a source term, Eq.~\eqref{eqf} describes free GWs. After quantization, their vacuum fluctuations in an expanding universe can be analyzed further. In this case, the exponential expansion of the universe leads quantum fluctuations of tensor modes to become a classical background for long-wavelength GWs. When quantum effects are neglected, GW are instead generated by the non-zero source term in Eq. (\ref{eqf}), where the source term  $\Pi^{TT}_{ij}$ denotes the transverse-traceless part of the anisotropic stress tensor $\Pi_{ij}$, 
\begin{eqnarray}
	\partial_{i} \Pi^{TT}_{ij}&=&\Pi^{TT}_{ii}=0,
\end{eqnarray}
and
\begin{eqnarray}
 \Pi_{ij}&=&T_{ij}-\langle p \rangle g_{ij},
\end{eqnarray}
where $T_{ij}$ refers to the energy momentum tensor and symbol $\langle p \rangle$ denotes the homogeneous pressure of the background.

Several methods can be used to solve Eq. (\ref{eqf}), including the use of Green functions in configuration space. On the other hand, in the harmonic gauge of Minkowski spacetime, the solution can be expressed in the wave zone approximation via a double Fourier transform (both time and space) of the stress-energy tensor~\cite{Khlebnikov:1997di, Easther:2006gt}. In this work, for the convenience of numerical calculation, such as preheating with extended sources or continuous media in an expanding universe, we work in Fourier space:

\begin{eqnarray}
	f(\bf k)&=&\frac{1}{(2\pi)^{3/2}}\int d^{3}{\bf x}e^{i{\bf k \cdot \bf x}}f(\bf x).
\end{eqnarray}
Accordingly, the wave function Eq. (\ref{eqf}) in Fourier space can be expressed as

\begin{eqnarray}
\ddot{h}_{ij}({t,\mathbf{k}})+ 3H \dot{h}_{ij}(t,{\mathbf{k}})+\frac{k^{2}}{a^{2}}h_{ij}(t,{\mathbf{k}})&=&\frac{16\pi G}{a^2} \Pi^{TT}_{ij}({t, \mathbf{k}}),\nonumber\\
\end{eqnarray}
the comoving wave-number is : $k=|{\mathbf{k}}|$. In the momentum space, the transverse-traceless part of $\Pi_{ij}$ is obtained by applying a projection operator~\cite{Garcia-Bellido:2007fiu}
\begin{eqnarray}
	\Pi^{TT}_{ij}({\bf k})&=&\big [ P_{il}(\hat{{\bf k}})P_{jm}(\hat{\bf k})-\frac{1}{2}P_{ij}(\hat{\bf k})P_{lm}(\hat{\bf k})\big ] \Pi_{lm}({\bf k}),\nonumber\\
\end{eqnarray}
where $\hat{\mathbf{k}}={\mathbf{k}}/k$ and $P_{ij}({\hat{\bf{k}}})=\delta_{ij}-{\hat{k}_i \hat{k}_j}$ is the projection operator onto the subspace orthogonal to the vector $\mathbf{k}$, satisfying $P_{ij} k_i = 0$ and $P_{ij} P_{jl} = P_{il}$. Hence, this leads directly to
\begin{eqnarray}
k_i\Pi^{TT}_{ij}&=&\Pi^{TT}_{ii}=0.
\end{eqnarray}

Based on the relation of the projection operator derived above, an arbitrary tensor in momentum space satisfies the following evolution relation:
\begin{eqnarray}
h_{ij}(t,{\bf k})&=&\Lambda_{ij,lm} u_{lm}({t,\mathbf{k}}).
\end{eqnarray}
Therefore Eq. (\ref{eqf}) can be rewritten as
\begin{eqnarray}
\ddot{u}_{ij}+3H \dot{u}_{ij}+\frac{k^2}{a^2} u_{ij}&=& \frac{16\pi G}{a^2} \Pi^{eff}_{ij}.
\end{eqnarray}
where $\Pi^{eff}_{ij}$ denotes the effective anisotropic stress tensor encapsulating the non-vanishing $\rm TT$ components of $\Pi_{ij}$. For the case of real scalar fields, $\Pi^{eff}_{ij}$ is explicitly given by
\begin{eqnarray}
\Pi^{eff}_{ij}&=&\partial_{i}\phi_a \partial_{j}\phi_a,
\end{eqnarray}
with $\phi_a$ representing the collection of real scalar fields indexed by $a=1,2..$.
Finally, the energy density spectrum of GW is expressed as
\begin{eqnarray}
\Omega_{\rm gw}(k) &=& \frac{1}{\rho_c}\frac{d\rho_{\rm gw}}{d\ln k} \nonumber\\
&=& \frac{k^3}{64\pi^{3} G V \rho_c} \int \frac{d\Omega_k}{4\pi} \, \dot{h}_{ij}(\mathbf{k}) \dot{h}^*_{ij}(\mathbf{k}).
\end{eqnarray}

\subsection{GW spectrum and observations}

\begin{figure*}[htbp]
    \centering
    \begin{subfigure}[b]{0.48\textwidth}
        \includegraphics[width=\linewidth]{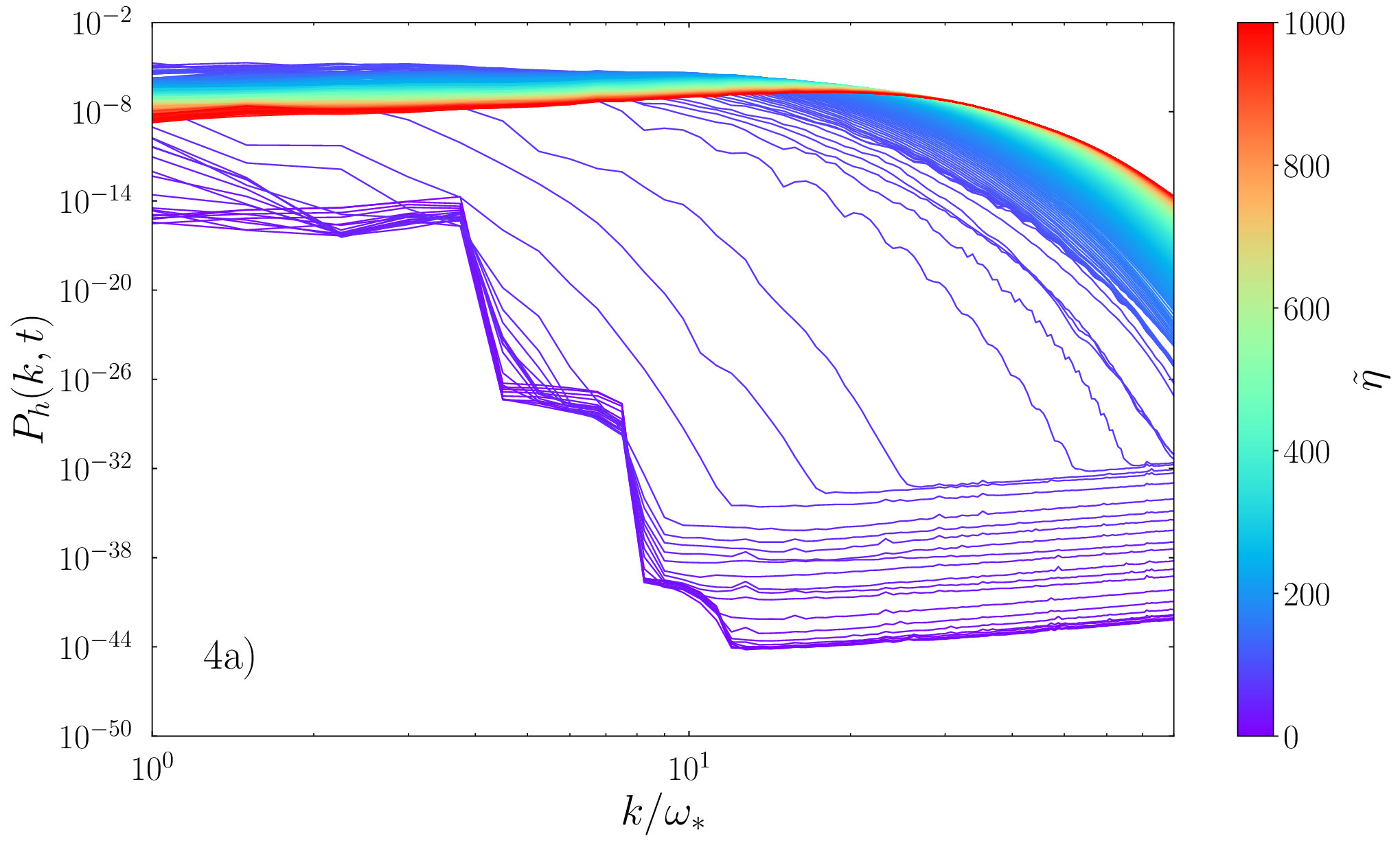}
        \caption{}
        \label{fig:bottom-left}
    \end{subfigure}
    \hfill
    \begin{subfigure}[b]{0.48\textwidth}
        \includegraphics[width=\linewidth]{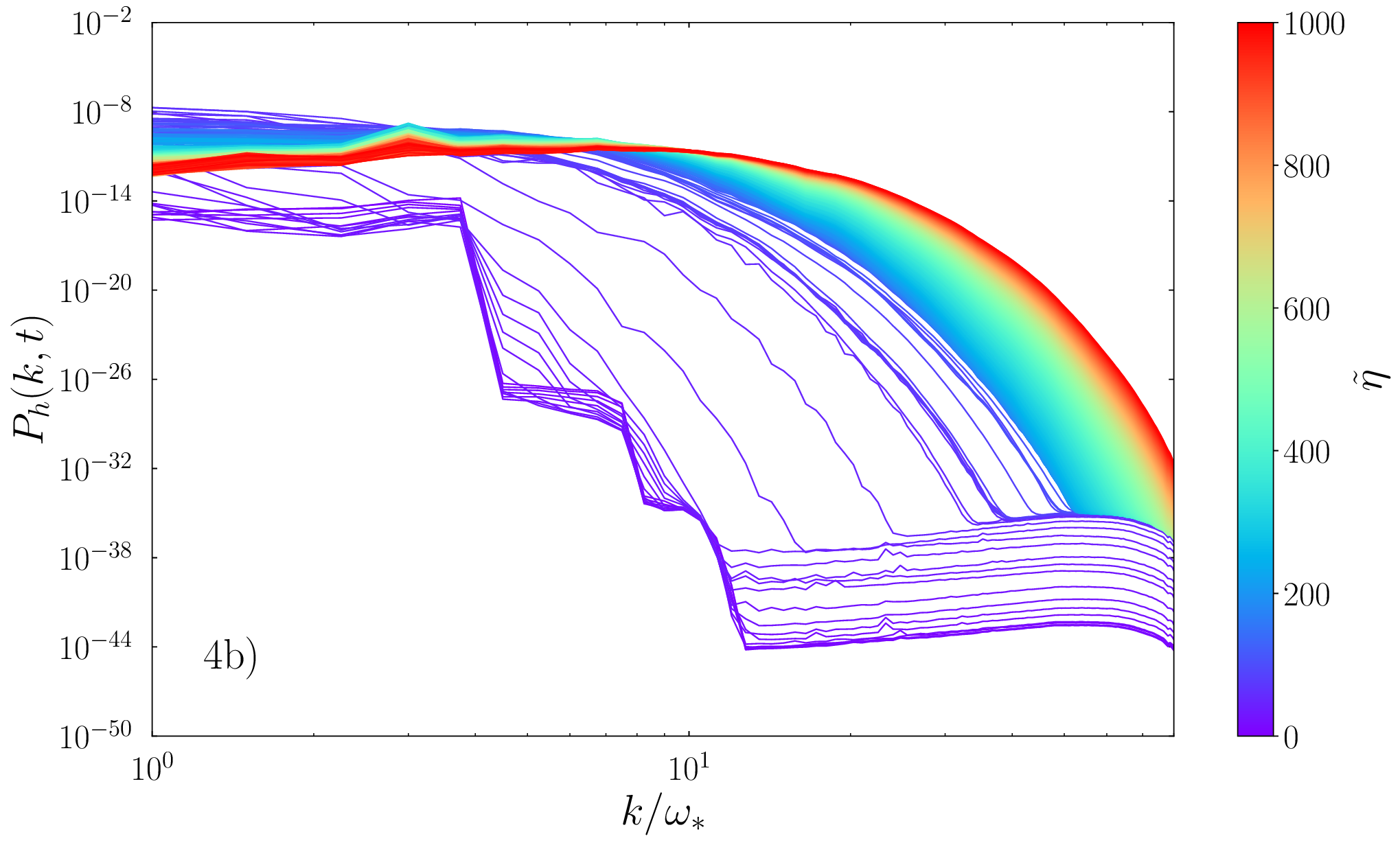}
        \caption{}
        \label{fig:bottom-right}
    \end{subfigure}
    \caption{The evolution of the Higgs field power spectrum $P_{h}(k,t)$ as a function of the wave number, where $\tilde{\eta}$ is the program variable time. The left and right panels correspond to $q_h=10^{2}$ and $q_h=10^5$, respectively.}
    \label{figmesh}
\end{figure*}

In the following analysis of the Higgs power spectrum and GW production, we adopt $q_h=10^2$, $q_{\phi h}=10^{2}$, and $q_\epsilon=10^{-6}$ as a representative benchmark point within the efficient-preheating region identified in Sec.~\ref{sec24}. We first present the evolution of the Higgs field power spectrum in Fig.~\ref{figmesh}. At early times, the spectrum is initially dominated by vacuum-like fluctuations. As the program time $\tilde{\eta}$ increases, the Higgs power spectrum is rapidly amplified, indicating the efficient excitation of Higgs field modes during the preheating stage. At later times, the spectrum gradually approaches saturation, and both its overall amplitude and shape become nearly stable. The enhancement of the Higgs power spectrum reflects the growth of Higgs field fluctuations and is consistent with an efficient transfer of energy from the inflaton field to the Higgs sector. This process is accompanied by the development of field inhomogeneities and an increase in gradient energy, which are characteristic features of the nonlinear stage of preheating. As the resonance develops further, backreaction and rescattering effects progressively suppress the growth of fluctuations, and the spectrum eventually saturates once the resonance becomes inefficient.

A clear dependence on the resonance parameter $q_h$ is observed. For $q_h=10^{2}$, the Higgs power spectrum is amplified more strongly, reaches a higher late-time amplitude, and retains larger power over most of the displayed wave-number range. This suggests that lower or moderate values of $q_h$ leads to a more efficient growth of Higgs fluctuations and is therefore consistent with a more effective transfer of energy from the inflaton field to the Higgs field. In contrast, for $q_h=10^{5}$, the amplification is weaker, and the late-time spectrum decreases more rapidly toward larger wave numbers, indicating that the excitation of Higgs modes is less efficient in this case.

\begin{figure}[htbp]
\centering
\includegraphics[width=0.48\textwidth]{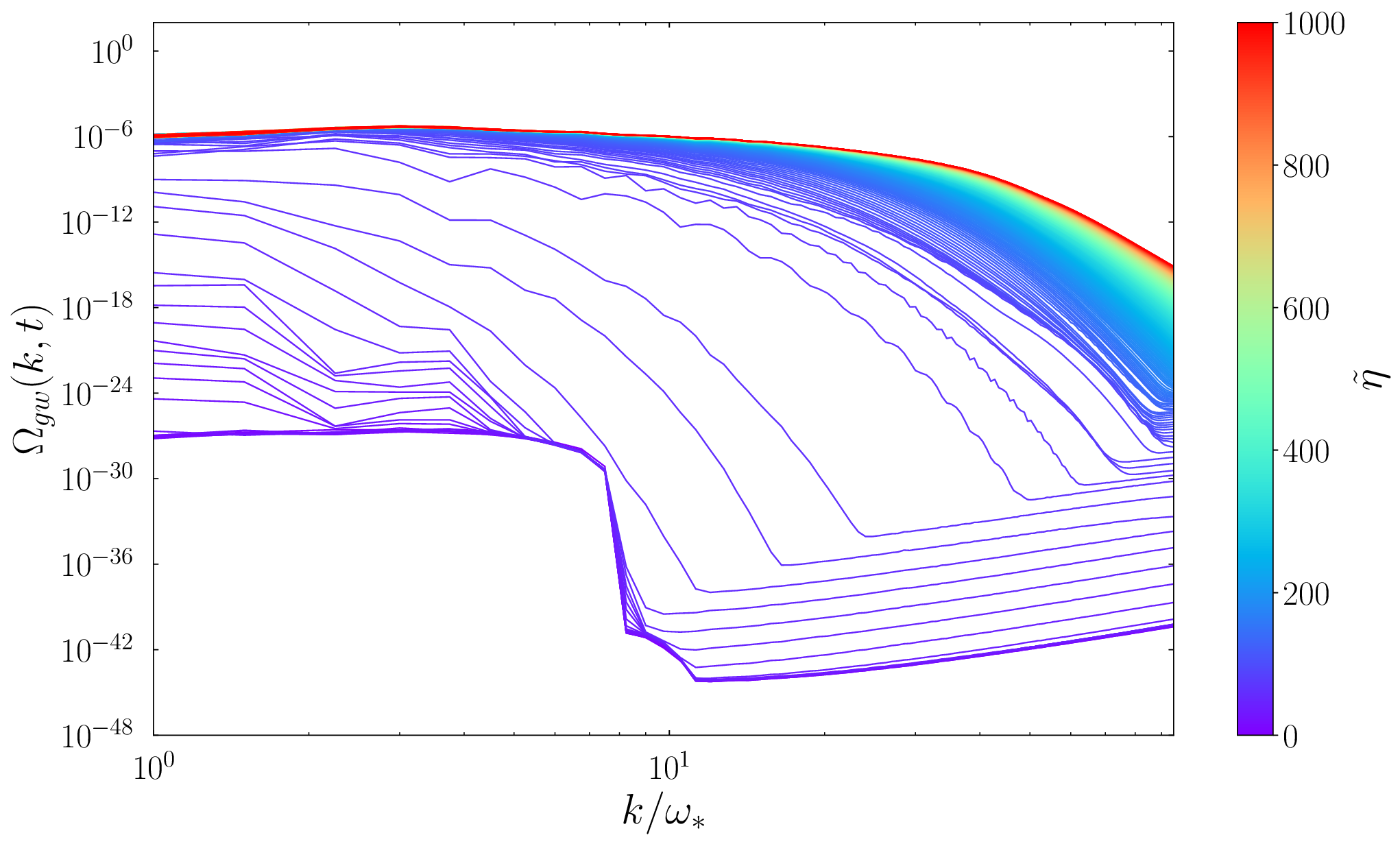}
\caption{Time evolution of the GW energy density spectrum \(\Omega_{\rm gw}(k,t)\). The color bar denotes the program time $\tilde\eta$. The benchmark parameters are $\lambda_{\phi}=10^{-13}$, $q_h=10^{2}$, $q_{\phi h}=10^{2}$, $q_\epsilon=10^{-6}$.}
\label{figspectrum}
\end{figure}

The behavior of the GW spectrum shown in Fig.~\ref{figspectrum} can be understood as a consequence of the preceding growth of the Higgs-field fluctuations. As the Higgs field becomes increasingly inhomogeneous, its spatial gradients enhance the anisotropic stress, which acts as the source of gravitational waves. Since the source term is quadratic in field gradients, the momentum distribution of Higgs fluctuations directly affects both the amplitude and the spectral shape of the resulting GW background. Fig. \ref{figspectrum} illustrates the evolution of the GW energy density spectrum $\Omega_{\rm gw}(k,t)$ obtained from the lattice simulation. The evolution can be divided into three distinct stages. Initially (purple curves), the GW spectrum remains strongly suppressed but is rapidly amplified as a consequence of the resonant growth of the Higgs field. In this stage, the spectrum exhibits a pronounced cutoff-like feature around the characteristic scale $k/\omega_* \sim 9$, which is consistent with the dominant momentum scale of the source fluctuations. Then, as the system enters the nonlinear evolution stage (blue to green curve), the rescattering effect becomes important, and the energy is redistributed toward higher-momentum modes. This leads to a significant broadening of the spectrum to the ultraviolet (UV) region and softens the sharp early-time spectral structure. In the final stage (orange and red curves), as the system approaches a quasi-stationary state, the growth of the gravitational wave signal reaches saturation. The final spectrum develops a broad peak with an amplitude of order $\Omega_{\rm gw} \sim 10^{-6}$ and exhibits a clear suppression at high frequencies. Figs.~\ref{figmesh} and \ref{figspectrum} show a consistent physical picture: amplification of Higgs-field fluctuations leads to stronger anisotropic stress and consequently becomes a source of gravitational waves.

\begin{figure}[htbp]
\centering
\includegraphics[width=0.48\textwidth]{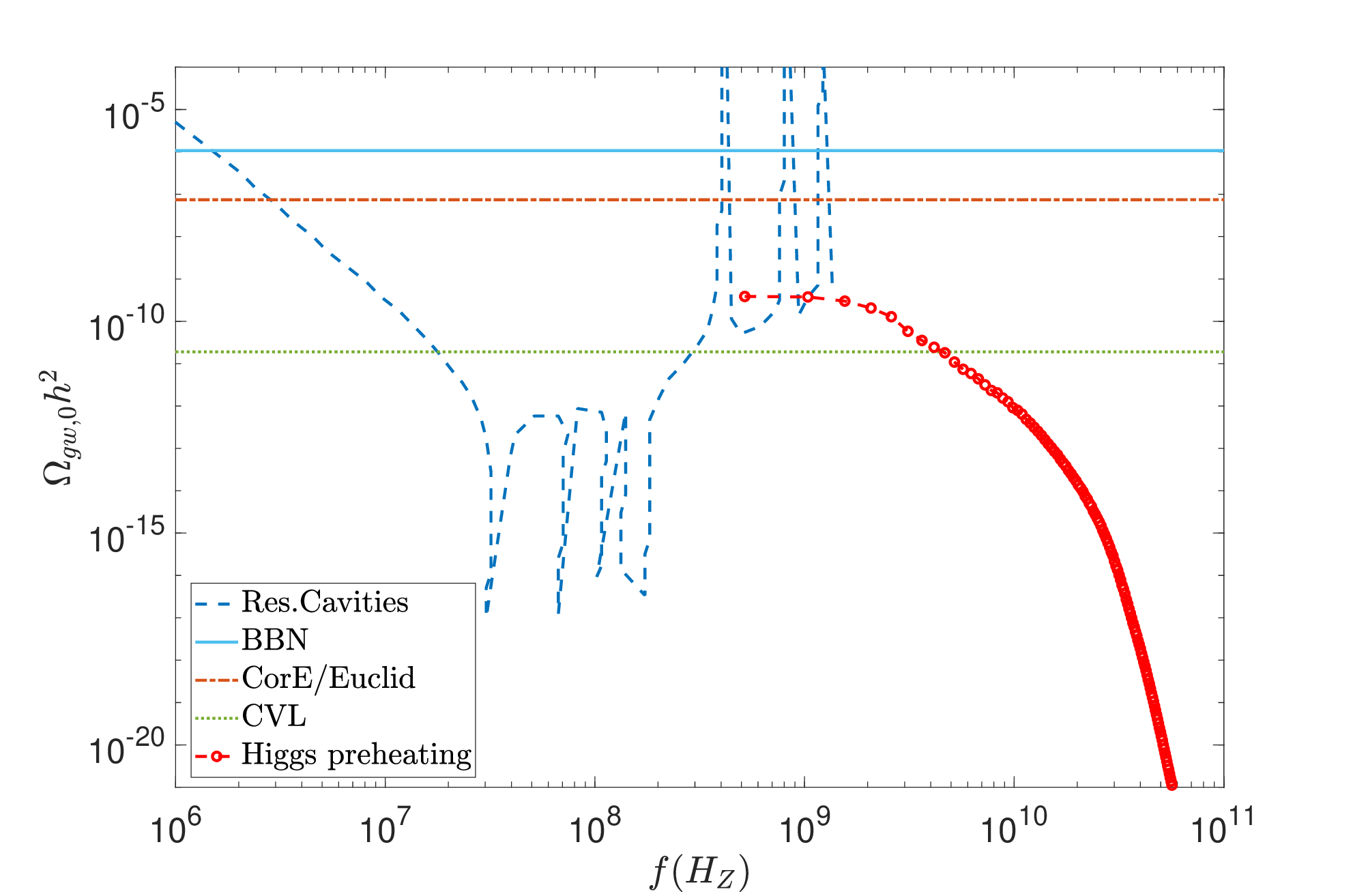}
\caption{The present-day GW energy density spectrum $\Omega_{\rm gw,0}h^2$ predicted in this work, compared with the projected sensitivity of resonant cavity detectors and indirect cosmological constraints. The benchmark parameters are $\lambda_{\phi}=10^{-13}$, $q_h=10^2$, $q_{\phi h}=10^{2}$, and $q_\epsilon=10^{-6}$.}
\label{fig8}
\end{figure}

The GW energy spectrum shown in Fig.~\ref{fig8} has been transformed to present-day observables, $\Omega_{\rm gw,0} h^2$, using the standard conversion relations given in Ref.~\cite{Dufaux:2008dn}, with the detailed transition relationship can be found in Ref.~\cite{Garcia-Bellido:2007fiu}. Fig.~\ref{fig8} further compares the predicted GW energy density spectrum from our model with the projected sensitivity of resonant cavity (Res.Cavities) detectors~\cite{Herman:2020wao,Herman:2022fau}, as well as with indirect constraints on the high-frequency GW background from future cosmological probes, including COrE/Euclid and the cosmic-variance-limited (CVL) case~\cite{EUCLID:2011zbd,COrE:2011bfs,Ben-Dayan:2019gll}. For completeness, the BBN bound is also shown. Our results indicate that at frequencies near $10^{9}$ Hz, the corresponding GW background partially overlaps with the projected sensitivity of resonant cavity detectors, suggesting that the high-frequency GW signal generated during Higgs preheating may be accessible to such experiments.

\section{Conclusion}
\label{sec4}
In this work, we have performed a systematic investigation of Higgs preheating in an inflaton-Higgs model with both trilinear and quartic interactions, using nonlinear lattice simulations. We have shown that the efficiency of preheating is controlled by three dimensionless parameters: the inflaton-Higgs coupling $q_{\phi h}$, the Higgs self-coupling $q_h$, and the mass-scale parameter $q_\epsilon$. Our numerical results indicate that efficient preheating requires $q_h < 10^{3}$, $q_{\phi h} > 10$, and $q_\epsilon < 10^{-5}$ when $\lambda_\phi=10^{-13}$. Beyond this range, either the resonance is too weak to transfer energy efficiently, or strong self-interactions induce early backreaction that suppresses particle production. We further find that, although the mean Higgs field remains small throughout the evolution, parametric resonance efficiently amplifies its inhomogeneous fluctuations, leading to significant anisotropic stress. In turn, this sources a gravitational wave background. 

The resulting gravitational-wave spectrum exhibits a characteristic three-stage evolution: an initial exponential growth during resonance, spectral broadening due to rescattering, and eventual saturation in the nonlinear regime. The final spectrum develops a broad peak with amplitude $\Omega_{\rm gw} \sim 10^{-6}$ at production. After redshifting to the present epoch, the gravitational wave peaks at frequencies around $f \sim 10^{9}\,\mathrm{Hz}$, with a present-day amplitude $\Omega_{\rm GW,0} h^2 \sim 10^{-10}$, lying within the projected sensitivity range of resonant cavity detectors while remaining compatible with current cosmological constraints. It is worth noting that the requirement $q_h \lesssim 10^3$ corresponds to a relatively small effective Higgs self-coupling for the chosen inflaton coupling. This may indicate that the Higgs quartic coupling is suppressed at the relevant energy scale, or that the scalar field should be interpreted as an effective Higgs-like degree of freedom.

\section*{Acknowledgments}
This work was supported by the Natural Science Foundation of China under Grant No. 12305091, by Sichuan Science and Technology Program No.2026NSFSC0758 and No.2024NSFSC1367, by the Research Fund for the Doctoral Program of the Southwest University of Science and Technology under Contract No.24zx7117 and No.23zx7122, by the Natural Science Foundation of Chongqing under Grant No.KJQN202503423, by Chongqing Natural Science Foundation project under Grant No.CSTB2022NSCQ-MSX0432, by Science and Technology Research Project of Chongqing Education Commission under Grant No.KJQN202200621, and by Chongqing Human Resources and Social Security Administration Program under Grants No.D63012022005.

\end{document}